# A Secure Medical Record Sharing Scheme Based on Blockchain and Two-fold Encryption

Md. Ahsan Habib
*Department of Computer Science and Engineering (CSE)*
*Khulna University of Engineering & Technology (KUET)*
*Khulna-9203, Bangladesh*
mahabib@cse.kuet.ac.bd

Kazi Md. Rokibul Alam
*Department of Computer Science and Engineering (CSE)*
*Khulna University of Engineering & Technology (KUET)*
*Khulna-9203, Bangladesh*
rokib@cse.kuet.ac.bd

Yasuhiko Morimoto
*Graduate School of Advanced Science and Engineering*
*Hiroshima University*
*Higashi-Hiroshima 739-8521, Japan*
morimo@hiroshima-u.ac.jp

*Abstract*–Usually, a medical record (MR) contains the patients' disease-oriented sensitive information. In addition, the MR needs to be shared among different bodies, *e.g.*, diagnostic centres, hospitals, physicians, etc. Hence, retaining the privacy and integrity of MR is crucial. A blockchain based secure MR sharing system can manage these aspects properly. This paper proposes a blockchain based electronic (e-) MR sharing scheme that (*i*) considers the medical image and the text as the input, (*ii*) enriches the data privacy through a two-fold encryption mechanism consisting of an asymmetric cryptosystem and the dynamic DNA encoding, (*iii*) assures data integrity by storing the encrypted e-MR in the distinct block designated for each user in the blockchain, and (*iv*) eventually, enables authorized entities to regain the e-MR through decryption. Preliminary evaluations, analyses, comparisons with state-of-the-art works, etc., imply the efficacy of the proposed scheme.

## I. INTRODUCTION

Generally, human beings are susceptible to disease, and to be cured, they need to visit health service-related bodies. As an example, a US citizen visits 18.7 different physicians on average and keeps 19 separate medical records (MRs) in its lifespan [1]. This figure is presumed to be higher in a developing country like Bangladesh. Also, the patient's MR may exist in diverse domains, and to serve better medical treatment, very often the MR needs to be shared among multiple bodies [2, 3]. However, sharing MR is challenging due to various constraints, *e.g.*, the data format, unreliable diffusion, interoperability, confidentiality violation, data integrity disruption [5], etc.

Preserving and exchanging the patients' MR through the conventional system is problematic and faces numerous hassles. Hence, electronic (e-) MR is replacing paper-based MR rapidly [6] where e-MR is usually stored in the local dedicated server of a healthcare service provider. However, a local server usually experiences numerous difficulties, *e.g.*, diverse malicious security attacks, single-point failure [7], etc., which may lead to forgoing data privacy. For instance, millions of e-MR had been compromised and healthcare databases had lost nearly $30 billion over the last two decades [2]. In addition, when a patient needs to share its e-MR with a healthcare service provider, it undergoes an inefficient and manual consent process [7]. Alongside, hackers had sold patients' data at up to 20 times higher prices than banking data [2].

This paper develops a blockchain based secure e-MR sharing scheme where a patient has sufficient control over its e-MR including the ability to know who has accessed it. Here, the major contributions are as follows.

*i)* Consideration of the medical image along with the textual data as the input.

*ii)* Enrichment of data privacy via a two-fold encryption mechanism consisting of an asymmetric cryptosystem and dynamic DNA encoding. Where at first, it encrypts the plain e-MR by an asymmetric cryptosystem. Then, applies the dynamic DNA encoding over the cipher e-MR to enrich the degree of confusion, *i.e.*, data privacy.

*iii)* Assurance of data integrity by storing each cipher e-MR in a distinct block having an index number in the blockchain. While storing, it returns the index number of the corresponding block to the concerned user for further usage.

*iv)* Enablement of authorized entities to regain the e-MR through decryption while maintaining the patients' anonymity.

The remaining parts of the paper are arranged as follows. Section II studies the related works. Section III describes the required building blocks. Section IV illustrates the proposed secure e-MR sharing scheme. Section V explains the preliminary evaluations, and Section VI concludes the paper.

## II. RELATED WORKS

As discussed in the previous section, e-MR contains patients' confidential information which needs to be shared among multiple bodies. In order to retain data privacy, data integrity, availability, interoperability, reliability, etc., for e-MR sharing system many schemes already have been proposed, *e.g.*, MediChain [5], HDG [8], MediBchain [9], MeDShare [10], MedBlock [11], DPS [12], SEMRES [13], MedChain [14], EHRChain [15], etc. Usually, these works adopt single-fold encryption to retain data privacy and store the plain e-MR metadata or encrypted e-MR over the blockchain.

The works proposed in [5], [8], [12], and [16] considered both medical images and textual data as input. The scheme in [12] used a symmetric key cryptosystem named Advanced Encryption Standard (AES) to encrypt the e-MR while to encrypt the AES key it used an asymmetric key cryptosystem named the Elliptic Curve Cryptography (ECC). In contrast, the schemes in [5], [8], and [16] did not explicitly mention their adopted cryptosystem. Alongside, the system in [8] used the blockchain to store the e-MR, whereas the scheme in [16] only stored the data-accessing-related information of the e-MR. The systems in [5] and [12] exploited location-related information in the blockchain.

The systems in [9] and [17] considered only textual data as input, where the approach [9] employed ECC to encrypt e-MR. But, to secure the e-MR and AES key, the scheme in [17] used the AES and threshold cryptosystem, respectively. The system in [9] used the permissioned blockchain to store the e-MR, whereas the work in [17] kept the metadata of the e-MR in the blockchain. A different work proposed in [18] took into account only the medical image to store on the local server. It employed an asymmetric-key cryptosystem to secure the URL of the individual file that is stored on the blockchain server. The system developed in [19] utilized a private and a consortium blockchain to preserve the e-MR and its index, respectively. Here, to encrypt the e-MR, it also used a public key cryptosystem.

Some other works proposed in [6], [7], and [13] also used the blockchain and employed the symmetric cryptosystem to secure the e-MR data. Here, schemes in [7] and [13] used an asymmetric cryptosystem to encrypt the symmetric key. The work in [6] stored the e-MR in the blockchain and to manage each block it replaced the traditional Merkle tree with an improved convolutional one to reduce the tree layers, nodes, hash calculations, etc. The scheme in [7] kept the e-MR and its metadata in the distributed file system and the blockchain server, respectively. In contrast, the system in [13] used the blockchain to store the hash value of the e-MR. However, none of these works specified their exact input.

The above analyses infer that to secure the e-MR data, huge state-of-the-art works rely on single-fold symmetric or asymmetric cryptosystem. Where a symmetric system like AES is vulnerable due to the side-channel attack, co-relation scan attack [20], etc. Likewise, known-plaintext, chosen-plaintext, ciphertext-only, etc., attacks can mount over the asymmetric cryptosystem [21]. Hence, to enrich data privacy this paper adopts a two-fold encryption technique capable enough to safeguard the sensitive e-MR data. In addition, rather than solely the textual data, it treats the visual and the textual data as input. Moreover, it deploys blockchain to assure data integrity and regains plain e-MR data while retaining the data owner's anonymity. The preliminary assessment shows that the proposed secure e-MR sharing scheme is proficient enough to retain colossal data privacy, data integrity, availability, authenticity, etc., criteria.

III. PRELIMINARIES

The major building blocks required to develop the proposed scheme are described below. In addition, it adopts an anonymous authentication mechanism, *i.e.*, deploys the anonymous credential proposed in [4] for each patient. Thus, while communicating with other entities, the patient can reveal its identity anonymously instead of the real identity.

*A. Blockchain Storage*

Blockchain is a decentralized, distributed, and tamper-proof data storage that confirms data integrity [5]. Nowadays it is widely used in different applications, *e.g.*, healthcare, financial services, public administration, supply chain management [22], etc. The data stored in the blocks of the blockchain are linearly and cryptographically linked together to form a chain. Each block contains a cryptographic hash pointer of the previous block, a timestamp, and transaction data. The hash pointer linked with the previous block gives the immutability property of the blockchain. Here, new blocks are added only when majority of the nodes consent by validating all transaction data. Since it appends new blocks continuously, it keeps growing. Fig. 1 depicts its general architecture.

Each block in the blockchain consists of two parts, *i.e.*, a block header and a block body. The block header contains the metadata that typically includes the block ID, the hash of previous block, the number of transactions, nonce, Merkle root, timestamp, etc. The block body holds each transaction data. As shown in Fig. 1, the Merkle tree, a binary hash tree is used to create the Merkle root that is stored in the block header. Here $D_i$ represents transaction data and $H_i$ denotes the cryptographic hash of the transaction $D_i$. SHA–256 is a prominent hash function [23] used in the blockchain domain.

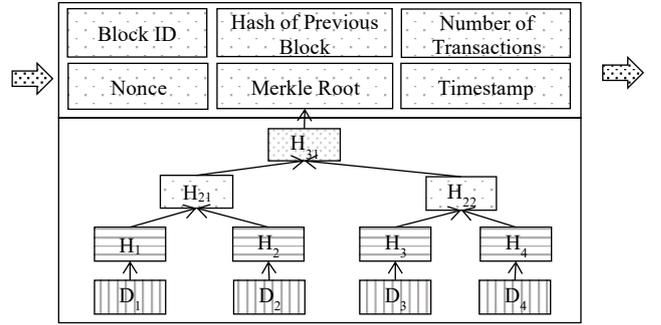

Fig. 1. The general architecture of blockchain.

*B. Dynamic DNA Encoding*

The goal of dynamic DNA encoding is to increase the confusion level of the ciphertext. For this purpose, each pair of successive cipher chunks are joined together. Different from [22], the DNA bases merge any two consecutive cipher chunks dynamically through the following equations.

$$R = log_b (x) \quad (1)$$

$$S = (R \times N) \ (mod \ Q) \quad (2)$$

Here, $b$ and $x$ specify the base and the random integer, respectively, whereas the values of $N$ and $Q$ are 10000 and 100, respectively. The $S$ dummy DNA bases are added in between every two consecutive chunks. The value of $x$ is incremented by 1 for calculating the value of $R$ for the next two consecutive chunks. These DNA bases are picked from the first chunk or the second chunk of two consecutive chunks determined via the following equation.

$$w = (-1)^S \quad (3)$$

If $w$ is positive, then $S$ dummy bases will be picked from the first chunk. Otherwise, $S$ dummy DNA bases will be picked from the second chunk.

*C. Formation of Two-fold Encryption*

The proposed two-fold encryption mechanism comprises the ElGamal cryptosystem [25] and the dynamic DNA encoding. Firstly, the plain data is encrypted through an asymmetric cryptosystem, *e.g.*, ElGamal. Here, ElGamal is chosen because its key generation is based on the discrete logarithm problem that is difficult to solve. Besides, the ciphertext produced by the ElGamal encryption function is non-deterministic and it performs better while decryption [22]. Then, dummy DNA bases are picked and placed within the chunks of encrypted data by using the dynamic DNA encoding mechanism to enrich the data privacy. Fig. 2 illustrates the formation of this encryption process.

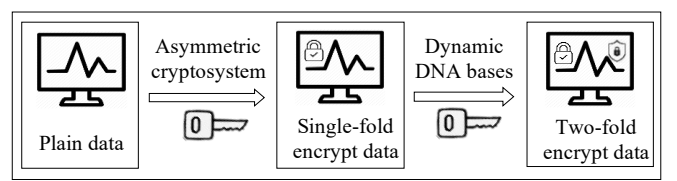

Fig. 2. The two-fold encryption process.

## IV. Proposed E-MR Sharing Scheme

This section first briefly describes the system model and provides an overview of the proposed e-MR sharing scheme which is depicted in Fig. 3. Then it describes the encryption, the blockchain storage, and the data query together with the decryption phases.

### A. System Model

The proposed scheme consists of five entities, *i.e.*, (*i*) $P_i$ ($i \geq 1$) patient, (*ii*) $Ph_j$ ($j \geq 1$) physician, (*iii*) $DC$ ($\geq 1$) diagnostic center, (*iv*) a blockchain storage, and (*v*) authorized third-party. Here, every entity possesses a unique identity. The major attribute of every entity are as below.

*Patient ($P_i$):* A person who needs medical treatment.

*Physician ($Ph_j$):* An individual who provides medical treatment to cure the patient's disease.

*Diagnostic Center (DC):* A pathological laboratory where a patient's clinical specimen is examined.

*Blockchain Storage (BS):* A decentralized, distributed, and tamper-proof data storage that ensures data integrity.

*Authorized Entity (AE):* An individual who is eligible to access the patients' health-related data to provide better service, analyses, etc.

### B. Overview of the Proposed Scheme

Let, for medical services a patient $P_i$ visits the physician $Ph_j$ to consult about its disease while the $P_i$ shares its public encryption key $Pub_P$ and other required parameters with the $Ph_j$. The $Ph_j$ asks for required medication and diagnosis that creates the $P_i$'s e-MR. Now using the $Pub_P$, $Ph_j$ encrypts the e-MR through the adopted two-fold encryption mechanism. Then the $Ph_j$ creates a block and forwards it to the blockchain network. Later on, while any authorized entity $AE$ requests the e-MR of $P_i$, the $P_i$ requests its block from the blockchain storage $BS$. Then the $BS$ returns the corresponding block to the $P_i$ and the $P_i$ decrypts the e-MR using its private key $Pri_P$ to obtain the plain e-MR. Further, the $P_i$ encrypts its e-MR using an asymmetric cryptosystem, namely ElGamal with different parameters using the public key $Pub_{AE}$ of $AE$ and sends it to $AE$. Then the $AE$ decrypts the cipher e-MR through its private key $Pri_{AE}$ to retrieve the plain e-MR.

### C. Encryption Phase

The physician $Ph_j$ encrypts the e-MR of the patient $P_i$ according to the following steps. Fig. 4 represents them.

*Step 1:* First, scan the textual data and the medical image (color) of the e-MR.

*Step 2:* Convert them separately into the corresponding ASCII value. Namely, for each character of the text, convert it into 03-digit ASCII integer. For the image, convert each pixel into its corresponding 03-channel ASCII value into 03-digit format. Now, these 03-channels are managed serially.

*Step 3:* Divide the ASCII integer data into equal-length chunks. If necessary, apply '0' padding at the leftmost of the leftmost chunk.

*Step 4:* Encrypt each chunk by using an asymmetric cryptosystem, *e.g.*, ElGamal encryption technique.

*Step 5:* Convert each encrypted chunk into its corresponding bits.

*Step 6:* Transform bits into corresponding DNA bases (*i.e.*, 00 = A, 01 = C, 10 = G, 11 = T).

*Step 7:* Add dummy DNA bases between every two consecutive chunks according to Section III-*C*. Finally, create a block of blockchain by using the encrypted chunks.

### D. Blockchain Storage Phase

After encryption, the $Ph_j$ creates a block with the encrypted e-MR, timestamp, and previous block hash and forwards it to the blockchain network. The $BS$ returns an index number $I_{BS}$ of the block to the $Ph_j$. Then the $Ph_j$ sends $I_{BS}$ to the $P_i$ for further accessing of the block.

### E. Data Query and Decryption Phase

When the $P_i$ wants to retrieve its e-MR, it anonymously sends an access request of the block containing the e-MR to the $BS$ providing the index number $I_{BS}$. Then, the $BS$ returns the corresponding block to the $P_i$ and the $P_i$ decrypts the cipher e-MR thru the following steps and Fig. 5 shows them.

*Step 1:* Scan the encrypted chunks of the block from the blockchain.

*Step 2:* Discard dummy DNA bases between every two consecutive chunks.

*Step 3:* Decode the DNA encoded chunks to retrieve the binary chunks.

*Step 4:* Convert the binary chunks into corresponding ASCII integer chunks (still it is in partial encrypted format).

*Step 5:* Decrypt the individual ASCII integer chunks by using the private key of the patient.

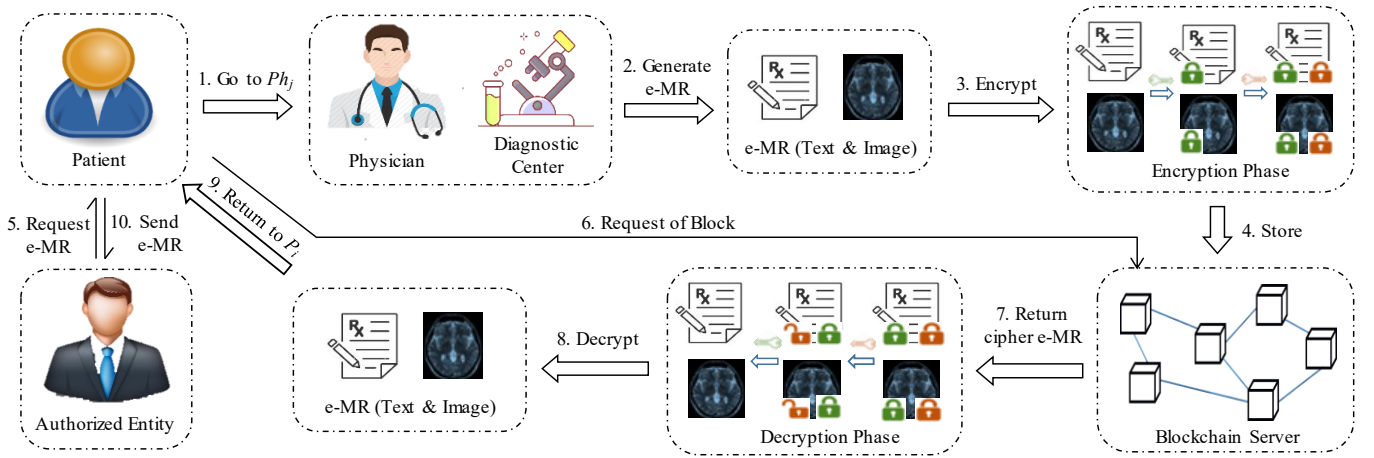

Fig. 3. An overview of the proposed scheme.

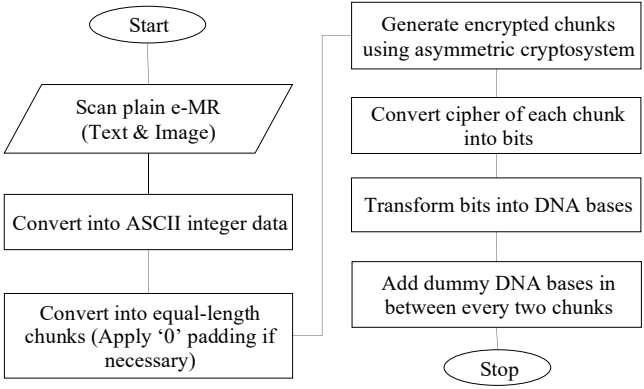

Fig. 4. Proposed encryption process.

*Step* 6: Discard all '0's from the left side of the left-most chunk, if necessary.

*Step* 7: Retrieve the e-MR (textual data and medical image separately) from the ASCII data.

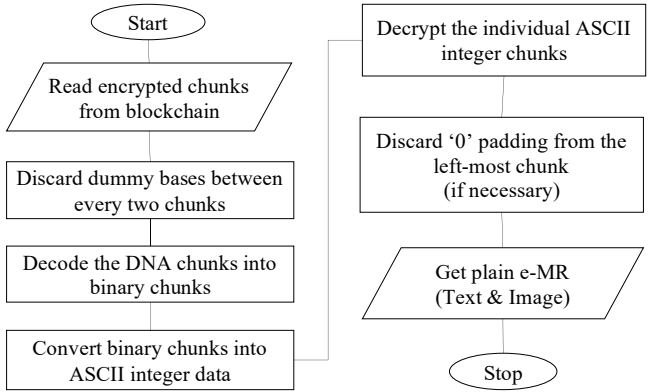

Fig. 5. Proposed decryption process.

## V. EVALUATION OF THE SCHEME

### A. Experimental Setup

A prototype of the proposed scheme was developed under the environment of Intel[(R)] Core™ i5-10500 CPU @3.10GHz 64-bit processor with 12 GB RAM running on Windows 10 operating system. It was developed in Visual Studio Code 2019 for coding purposes. The ElGamal encryption system is adopted as an asymmetric cryptosystem and used a 1024-bit key for encryption and decryption operations. Byte size matters calculator [23] was used to measure the text size. Here, the parameters related to the dynamic DNA encoding, *i.e.*, to form two-fold encryption phase were set as $b=3$ and $x = 10$ where the other parameters were already specified in Section III-*C*.

### B. The output of the Encryption Phase

Considering the plaintext 'Patient Name: Alice' and a color chest X-ray image with a size of 300×250, based on the steps of Section IV-*C*, Table I presents the output of encryption phase.

### C. The output of the Decryption Phase

Based on the steps of Section IV-*E* (and for the data of Section V-*B*), Table II shows the output of decryption phase.

### D. Experimental Results and Comparisons

The proposed e-MR sharing scheme was implemented with different data sizes by employing the developed two-fold encryption mechanism. Here, three sets of e-MR were chosen as the input, (*i*) 100KB textual data and a 200×200 color medical image (29KB), (*ii*) 300KB textual data and a 300×300 color medical image (263KB), and (*iii*) 500KB textual data and a 500×500 color medical image (732KB). Then the size of the plain e-MR together with the size of the cipher e-MR, is shown in Fig. 6. By observing the figure, it is seen that the size of the cipher e-MR is approximately 10 times greater than the size of the plain e-MR.

TABLE I. OUTPUT OF THE ENCRYPTION PHASE

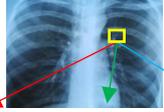

*denotes a portion of data; **dummy DNA bases are shown in bold italic

TABLE II. OUTPUT OF THE DECRYPTION PHASE

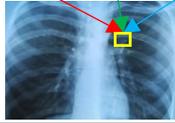

| Step | Operation | Output |
|---|---|---|
| *Step* 1 | Read final ciphertext | *,**ATCTGTTCTGTTACTCAATCCAACAACTTGGT*ATC*TGCTACGGGCGGCTATTTCT AGTACGATGAAACATTGCGCTTCCCAACCAACAATTGCGCTTGCTATTTCT*TTTCT* |
| *Step* 2 | Discard dummy bases | **Chunk 1:** *TGCAAGAACTGAAGGCCTCTAACCTCTACTTAGACCGTATGAGAAGAT <br> **Chunk 2:** *TAAGTCTAGCTTTCTTCCACACATGAGCTACCCAGGCGAATAAGCCAC |
| *Step* 3 | Decode into binary chunks | **Chunk 1:** *1100100001000000111100000101001011101110000010111011100011111100101 <br> **Chunk 2:** *1000010010100011110110111100001100011000111011001111110100010011001 |
| *Step* 4 | Convert into ASCII integer data | **Chunk 1:** *160196110354652489523396765104697193423769108011543972106147876233 <br> **Chunk 2:** *106871287860786913019408022543218433480150823923496926652762894356 |
| *Step* 5 | Decrypt ASCII integer chunks | **Chunk 1:** *0800971161051011101160320780971091010580320651081050991011600010240 <br> **Chunk 2:** *1861710171911390192391790291991390392391660162061230131830640644244 |
| *Step* 6 | Discard '0' padding from the left-most chunk | 0800971161051011101160320780971091010580320651081050991011- 6001024016601618617101719113901923917902919913903923916601620612301318306404 64244 |
|  |  | 0800971161051011101 160 166 171   010 016 017   240 186 191 <br> 6032078097109101 0580 139 179 139   019 029 039   239 199 239 <br> 32077114046032065108 166 123 064   016 013 064   206 183 244 <br> 105099101      Red channel     Green channel     Blue channel |
| *Step* 7 | Get plain e-MR (Text & Image) | **Text:** Patient Name: Alice    **Image:** |

*denotes a portion of data; **dummy DNA bases are shown in bold italic

Another experiment was done to exhibit the encryption and decryption time where the size of input e-MR sets remains the same as in Fig. 6. The output is displayed in Fig. 7. It shows that the time required for the encryption is nearly double than the time required for the decryption. The reason is that the proposed two-fold encryption mechanism at first employs the ElGamal cryptosystem that consists of two modular exponentiation operations in the encryption phase but only one alike operation in the decryption phase. Alongside, Fig. 8 depicts the time requirement to create a distinct private block while using the same inputs of Fig. 6 also and it excludes the e-MR encryption time.

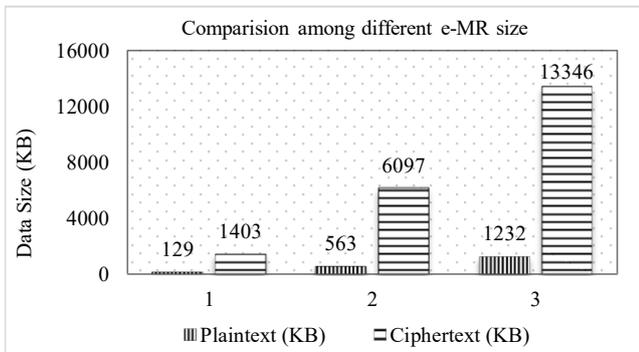

Fig. 6. A comparison between plain e-MR and cipher e-MR data-size for three distinct inputs.

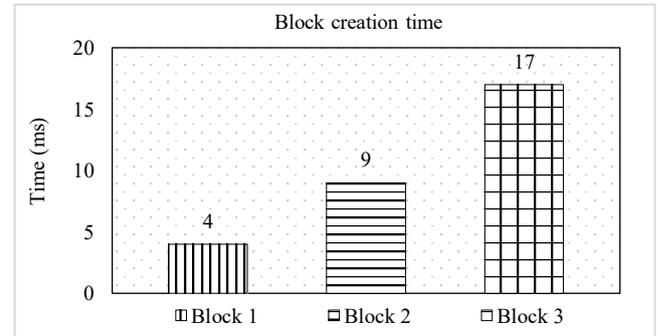

Fig. 8. Time required for block creation of three different cipher e-MR.

The proposed scheme is compared with [5]–[9], [12], [13], [16], [17] and [18] as shown in Table III. It shows that all the compared schemes uses a single-fold encryption and most of them use symmetric cryptosystem that is vulnerable for various attacks. Some of the works do not consider image data along with textual data while most of them do not maintain the data integrity. Thus, the table implies that the proposed scheme is better for storing and sharing e-MR and assures data privacy, data integrity, availability, etc.

*E. Security Analyses*

This section evaluates the security aspects, *i.e.*, data privacy, data integrity, level of encryption, etc., encompassed by the proposed scheme that are depicted in Table III. At the same time, the table includes a comparison with other works.

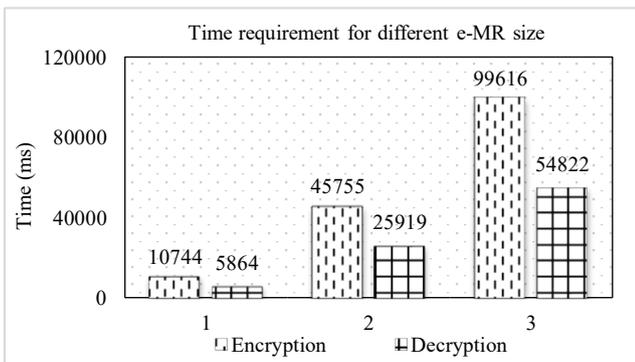

Fig. 7. A comparison between encryption and decryption time for three distinct e-MR inputs.

TABLE III. A COMPARISON WITH OTHER SCHEMES

| Scheme | Input type | | Encrypt e-MR asymmetrically | data integrity | Two-fold encryption |
|---|---|---|---|---|---|
| | Text | Image | | | |
| [6] | – | – | × | √ | × |
| [7] | – | – | × | × | × |
| [8] | √ | √ | – | √ | × |
| [9] | √ | × | √ | √ | × |
| [12] | √ | √ | × | × | × |
| [13] | – | – | × | × | × |
| [5, 16] | √ | √ | – | × | × |
| [17] | √ | × | – | × | × |
| [18] | × | √ | √ | √ | × |
| Proposed | √ | √ | √ | √ | √ |

– = not mentioned; √ = considered; × = not considered

*Data Privacy:* The proposed scheme offers privacy via ElGamal encryption and dynamic DNA encoding. But, as in Table III, most compared schemes use symmetric encryption

*Data Integrity:* As blockchain is an immutable ledger, data integrity of the cipher e-MR is assured by storing in a distinct block of the blockchain. Whereas, as in Table III, other almost schemes do not satisfy this criteria.

*Two-fold Encryption:* To enrich the level of confusion, the proposed scheme adopts two-fold encryption. However, as in Table III, the compared ones use single-fold encryption.

## VI. CONCLUSIONS

In this paper, the proposed blockchain based secure e-MR storing and sharing scheme maintains adequate data privacy, data integrity, availability, etc., about the patient's medical record. It considers both the textual data and the medical image as input. Herein, the two-fold encryption mechanism consisting of the asymmetric cryptosystem and the dynamic DNA encoding enriches data privacy. The storage of the patient's medical record over the blockchain assures data integrity. The preliminary assessment refers the efficacy of the proposed scheme. An upcoming plan of enhancement is to improve the required building blocks and the encryption and decryption phases, develop and incorporate a consensus mechanism for blockchain architecture, and implement the prototype of this scheme in a more realistic environment.